\documentclass[twocolumn,10pt]{asme2ej}         

\usepackage{graphicx}
\usepackage{caption}
\usepackage{subcaption}
\usepackage{xcolor}
\usepackage{amsmath}
\usepackage{amsfonts}
\usepackage{amssymb}

\newtheorem{remark}{Remark}

\title{\LARGE \bf
Data-driven design of steady-state feedforward inputs for nonlinear systems under partial measurement
}

\author{Sathya Aswath Govind Raju\thanks{Address all correspondence to this author. Co-first author.}
    \affiliation{Department of Mechanical Engineering\\
    University of Minnesota-Twin Cities\\
    Minneapolis, MN 55455, USA\\
    Email: govin091@umn.edu}
}

\author{Berk Altiner\thanks{Co-first author.}
    \affiliation{Department of Mechanical Engineering\\
    University of Minnesota-Twin Cities\\
    Minneapolis, MN 55455, USA\\
    Email: altin009@umn.edu}
}

\author{Zongxuan Sun
    \affiliation{Department of Mechanical Engineering\\
    University of Minnesota-Twin Cities\\
    Minneapolis, MN 55455, USA\\
    Email: zsun@umn.edu}
}

\author{Arunava Banerjee
    \affiliation{Department of Mechanical Engineering\\
    University of Minnesota-Twin Cities\\
    Minneapolis, MN 55455, USA\\
    Email: abanerje@umn.edu}
}

\author{Rajasree Sarkar
    \affiliation{Department of Mechanical Engineering\\
    University of Minnesota-Twin Cities\\
    Minneapolis, MN 55455, USA\\
    Email: sarka122@umn.edu}
}

\author{Kenneth Kim
    \affiliation{DEVCOM Army Research Laboratory\\
    Aberdeen Proving Ground\\
    Aberdeen, MD 21005, USA\\
    Email: kenneth.s.kim11.civ@army.mil}
}

\author{Chol-Bum Mike Kweon
    \affiliation{DEVCOM Army Research Laboratory\\
    Aberdeen Proving Ground\\
    Aberdeen, MD 21005, USA\\
    Email: chol-bum.mkweon2.civ@army.mil}
}

\begin{document}

\maketitle
% Removes page numbers for the final camera-ready copy submission
\thispagestyle{empty}
\pagestyle{empty}

\begin{abstract}
\textit{Designing trajectory tracking controllers for nonlinear systems remains a significant challenge, traditionally requiring precise mathematical models and complex analytical derivations. While the Internal Model Principle (IMP) provides a robust theoretical foundation for such problems, its application is often hindered by model uncertainty and the inherent complexity of nonlinear controller synthesis. This work proposes a practical data-driven control framework that bypasses the need for an explicit first-principles model by utilizing raw input-output data. By integrating fundamental results from IMP theory with nonlinear system analysis, the proposed approach improves design tractability. The framework's efficacy is validated through numerical simulations on two distinct nonlinear platforms: a mechanical load with a nonlinear friction term and an electrohydraulic actuator system.\\
Keywords: Data-driven control, Trajectory tracking, Nonlinear systems, Internal model principle, Partial measurement}
\end{abstract}

% \includegraphics{figs/grabs.pdf}

% \begin{highlights}
% \item A data-driven steady-state feedforward generator is designed for unknown nonlinear systems under partial measurement.
% \item The proposed framework operates in the frequency domain to exploit the harmonic structure of the steady-state input.
% \item The learned generator recovers the harmonic structure of the analytical regulator equations solution without requiring a plant model.
% \item The framework is validated on a mechanical system with nonlinear friction and an electrohydraulic actuator.
% \end{highlights}

% Keywords
% \begin{keywords}
% Data-driven control \sep Trajectory tracking \sep Nonlinear systems \sep Internal model principle \sep Partial measurement
% \end{keywords}

% \includegraphics{figs/grabs.pdf}


% Keywords
%\begin{keywords}
%quadrupole exciton \sep polariton \sep \WGM \sep \BEC
%\end{keywords}

\maketitle

\section{Introduction}
Trajectory tracking in the presence of disturbances, commonly referred to as output regulation, is one of the central problems in control theory. This problem arises in a wide range of practical applications, including motion control of electrohydraulic actuators \cite{sun2022robust}, tracking control of robot manipulators \cite{wu2025task}, and autonomous landing control of vertical takeoff and landing vehicles \cite{marconi2002autonomous}. The output regulation problem for linear time-invariant (LTI) systems has a mature theory and was extensively studied during the 1970s \cite{francis1976internal}, \cite{davison2003robust}. A major outcome of these works is the well-known \textit{internal model principle} (IMP) which plays a fundamental role in the solution of the output regulation problem. The extension of output regulation to general nonlinear systems was established in the seminal work of Byrnes and Isidori \cite{isidori1990output}, where solvability of the output regulation problem is characterized by the solvability of a set of nonlinear differential equations, commonly referred to as the regulator equations.

The key step in solving the output regulation problem is the design of internal models. Unlike LTI systems, in which the internal model can be constructed by embedding a copy of the exosystem within the controller, nonlinear systems introduce significant challenges in the design of the internal model. The main reason behind this challenge is the nonlinear distortion phenomenon that generates additional frequency components, including harmonics for single-frequency references and intermodulation components for multi-frequency references. In output regulation, internal models act as steady-state generators of the ideal feedforward control input that drives the regulation error to zero when the system is properly initialized. Consequently, the internal model must be designed to incorporate these distortions to achieve output regulation. 

The first step in internal model design is characterizing the steady-state feedforward input, including the harmonic content induced by the plant nonlinearity. Under specific structural assumptions on the steady-state feedforward input, such as polynomial or trigonometric-polynomial dependence on the exosystem states, canonical internal model structures are developed in \cite{byrnes1997structurally}, \cite{huang2004general}. For general nonlinear systems, however, exact characterization is generally intractable, and approximation methods are commonly employed. These include Taylor series expansions \cite{huang2002approximation,huang1994robust}, neural network approximations \cite{wang2001neural}, and, more recently, physics-informed neural networks \cite{mengozzi2025physics}. Data-driven approaches for nonlinear output regulation have been developed in \cite{wang2024nonparametric}, \cite{harry2025data} to address the aforementioned challenges as well as uncertainties in the exosystem.

Despite the advantages of these approaches, the model-based methods above require an explicit first-principles model, and the data-driven extensions remain tied to specific structural classes. In practice, constructing a representative first-principles model for complex nonlinear systems is itself a significant challenge. This has motivated the development of data-driven modeling and control approaches aimed at overcoming the limitations of classical model-based control. Among the vast body of literature on data-driven modeling and control, only a limited number of works have addressed the output regulation problem. For LTI systems with unknown dynamics, reinforcement-learning-based approaches have been proposed in \cite{chen2022robust, chen2019reinforcement}, and a Fundamental Lemma–based data-driven regulation framework has been developed in \cite{chen2025data}. For nonlinear systems, \cite{hu2025data} develops a data-driven version of the approximate regulation framework of \cite{astolfi2015approximate} for systems whose dynamics consist of known nonlinearities with unknown coefficients. Building on the same system class,  \cite{liu2025data} achieves asymptotic regulation by combining a data-driven passivity-based feedback that cancels the plant nonlinearities with an exosystem copy embedded in the controller. 

Despite the significant progress made by existing data-driven output regulation approaches, several fundamental limitations remain, particularly for nonlinear systems. Most existing methods are developed for a specific class of unknown systems whose dynamics can be expressed in terms of known nonlinear basis functions with unknown coefficients, and rely on the availability of full state measurements. These assumptions considerably simplify the system representation and controller synthesis, but may be restrictive in practice, where only partial output measurements are available and the underlying nonlinearities are complex or poorly characterized. Moreover, existing data-driven regulation frameworks typically assume that the internal model structure—either in the form of harmonic generators or exact exosystem dynamics—is known a priori and embedded directly into the controller. As a result, these approaches do not address the data-driven design of the steady-state feedforward generator under fully unknown plant dynamics and partial measurement. This gap motivates the development of data-driven design methods that operate directly on input–output data and produce the steady-state feedforward input required for trajectory tracking.

In this paper, we address the design of a data-driven steady-state feedforward generator for nonlinear systems with unknown dynamics under partial measurement. To handle the unknown plant dynamics, we adopt a data-driven approach and propose a method that produces the steady-state feedforward input corresponding to a given class of reference trajectories. Our approach consists of two main stages: data-driven modeling and steady-state feedforward generator design.

In the modeling stage, to address the partial measurement setting, we follow the framework presented in \cite{banerjee2025data}. Specifically, this framework leverages the history of inputs and measured outputs to learn a mapping that captures the input–output behavior of the unknown system.  However, time-domain histories can suffer from the curse of dimensionality, particularly at high sampling rates. To mitigate this issue, we transform the input-output histories into the frequency domain via the Fourier transform. This representation provides a lower-dimensional characterization of the system dynamics while retaining the essential information needed to capture its behavior. In addition, the frequency-domain representation provides insight into the nonlinear distortions induced by the system dynamics. This insight leads to an efficient parameterization of feedforward inputs that facilitates tracking of given reference trajectories. In the steady-state generator design stage, the learned frequency-domain model of the unknown nonlinear system is incorporated into a neural network training algorithm to construct a feedforward generator map. Specifically, this neural network mapping plays the role of the steady-state input generator associated with the IMP: it takes the amplitude and phase of the reference trajectory as inputs and generates a feedforward control signal in the frequency domain such that the tracking error is minimized.

Beyond enabling steady-state feedforward generator construction by leveraging input–output data, the proposed approach establishes a direct connection to classical internal model design. Polynomial solutions of the regulator equations produce harmonic content at integer multiples of the exosystem's fundamental frequency, with amplitudes determined by the plant nonlinearity \cite{huang2001remarks}. Classical internal model design reveals this harmonic structure by solving the regulator equations from a known plant model. In contrast, our approach treats the plant as a black box accessed only through partial measurements and recovers the dominant finite-bandwidth spectral structure through frequency-domain learning. The proposed framework can thus be interpreted as a data-driven realization of the frequency-domain representation of the steady-state feedforward control, obtained without explicit knowledge of the plant or solution of the regulator equations.

%This linkage stems from the fact that polynomial solutions of the regulator equations generate harmonic components of the fundamental tracking frequency, reflecting the underlying system dynamics. In classical internal model design, this harmonic structure is explicitly revealed through the solution of the regulator equations. In contrast, since the system is treated as a black box and only partial measurement data are available, we uncover this structure implicitly through data-driven analysis.

The rest of the paper is organized as follows. Section 2 provides the preliminaries. Section 3 outlines the proposed data-driven control framework. Numerical results are presented in Section 4. Finally, Section 5 concludes the paper and discusses future research directions.

\section{Preliminaries and Problem Definition}
In this section, we summarize the results that build the foundation of the proposed method. 

\subsection{Output Regulation for Nonlinear Systems}

Consider the general class of nonlinear systems
\begin{align}
\label{eq: plant}
    \dot{x}(t) &= f(x(t),u(t),v(t)), \\
    e(t) &= h(x(t),v(t)),
\end{align}
where $x(t)\in \mathbb{R}^{n_x}$ denotes the system state, 
$u(t) \in \mathbb{R}^{n_u}$ is the control input, and 
$e(t) \in \mathbb{R}^{n_e}$ is the regulated output. 
The exogenous signal $v(t) \in \mathbb{R}^{n_v}$ denotes both the reference to be tracked and the disturbance to be rejected, and it is generated by
\begin{equation}
\label{eq: exosystem}
    \dot{v}(t)=Sv(t), \quad v(0)=v_0,
\end{equation}
which is referred to as the exosystem.

The output regulation problem consists of designing a controller of the form
\begin{align}
\label{eq: controller}
    \dot{z}(t) &= f_c(z(t),e(t)), \\
    u(t) &= g_c(z(t)),
\end{align}
such that the following properties hold:

\begin{enumerate}
    \item The unforced closed-loop system
    \begin{align}
        \dot{x}&=f(x,g_c(z),0),\\
        \dot{z}&=f_c(z,h(x,0))
    \end{align}
    has a locally exponentially stable equilibrium at $(x,z)=(0,0)$.

    \item The forced closed-loop system
    \begin{align}
        \dot{x}&=f(x,g_c(z),v),\\
        \dot{z}&=f_c(z,h(x,v)),\\
        \dot{v}&=Sv,
    \end{align}
    satisfies
    \begin{align}
        \lim_{t \to \infty} e(t) = 0
    \end{align}
    for all initial conditions $(x(0),z(0),v(0))$ sufficiently close to $(0,0,0)$.
\end{enumerate}

Under standard stabilizability and detectability assumptions, solvability of the output regulation problem is characterized by the existence of smooth mappings $x=\pi(v)$ and $u=c(v)$ defining the invariant zero-error manifold corresponding to the steady-state behavior of the closed-loop system. These mappings are obtained as solutions of the regulator equations
\begin{align}
\label{eq: reg_eqn1}
    \frac{\partial \pi}{\partial v} Sv &= f(\pi(v),c(v),v), \\
    0 &= h(\pi(v),v).
\label{eq: reg_eqn2}
\end{align}

A controller that achieves output regulation must be capable of generating the steady-state input $c(v)$ driven by the exosystem \eqref{eq: exosystem}. In particular, a controller that achieves output regulation must contain a mechanism capable of reproducing the trajectories generated by
\begin{align}
\label{eq: generator1}
    \dot{v} &= Sv,\\
    u &= c(v).
\label{eq: generator2}
\end{align}

Constructing such a mechanism becomes systematic when the steady-state input $c(v)$ admits a polynomial representation in the exosystem states \cite{byrnes1997structurally, huang1994robust}. In particular, if
\begin{equation}
\label{eq: polynomial}
   c(v) = \sum_{i=1}^{n_p} A_i\, p_i(v),
\end{equation}
where $p_i(v)$ are monomials in the exosystem states, and $A_i$ are coefficients, then a finite-dimensional internal model can be systematically constructed. 

An important consequence of this polynomial representation is that, for a linear exosystem $\dot v = Sv$ with purely imaginary eigenvalues, the steady-state control input consists of a finite combination of harmonic components. The equivalence between the polynomial representation and the harmonic structure of the steady-state input is established in \cite{huang2001remarks}. Accordingly, the steady-state control input can be represented as a finite sum of complex exponentials
\begin{equation}
\label{eq: frequency}
   c(v(t)) = \sum_{i=0}^{n_f} C_i(v_0)\, e^{j\omega_i t},
\end{equation}
where the fundamental frequencies are determined by the exosystem dynamics, while higher-order harmonic components arise from nonlinear combinations induced by the polynomial structure of $c(v)$, which reflects the underlying plant dynamics. The coefficients $C_i(v_0)$ depend on the system parameters and the initial conditions of the exosystem, i.e., the amplitudes and phases of the exogenous signals.

While the formulation above describes the full closed-loop regulation problem, this paper focuses on the data-driven design of the steady-state feedforward input $c(v)$, which constitutes the steady-state component of any internal model-based controller.

\subsection{Problem definition}
In the standard model-based design reviewed above, several limitations arise. First, solving the regulator equations \eqref{eq: reg_eqn1}–\eqref{eq: reg_eqn2} requires an accurate structural description of the plant dynamics. Although the internal model principle provides robustness with respect to parameter variations, the correct functional form of the nonlinearities must be known \cite{bin2018chicken}. Missing structural terms, such as monomials in the state dynamics, may lead to the omission of the corresponding harmonic components in the steady-state input. Second, even when the structural model is available, solving the regulator equations is, in general, a highly challenging task for nonlinear systems.

Motivated by these limitations, we pose the following question:
\textit{Can the ideal steady-state control input be identified without relying on an explicit first-principles plant model?}

To address this question, we build our framework upon the equivalence results established in \cite{huang2001remarks} and present a data-driven framework in the frequency domain for steady-state feedforward generator design in the next section.

\section{Frequency-based Steady-State Generator Design Methodology}
The proposed data-driven framework consists of two stages: model generation and steady-state generator design. In the model generation stage, we motivate the use of a frequency-domain representation and discuss its advantages over time-domain formulations. In the steady-state generator design stage, we present a method that identifies the harmonic coefficients producing the feedforward input that minimizes tracking error.
\subsection{Model Generation}

In this paper, we consider the general class of nonlinear systems described in \eqref{eq: plant}, with the additional constraint that only partial state information is available through output measurements of the form
\begin{equation}
y(t) = g(x(t),u(t)),
\end{equation}
which is a common scenario in practice due to sensing limitations, cost considerations, and physical constraints.

In this setting, the state transition function $f$, the output function $g$, and the number of states are assumed to be unknown. The only available information about the system consists of input–output data collected from a series of experiments. Specifically, we assume access to a dataset
\begin{equation}
\label{eq:data}
\mathcal{D} = \{(U_i, Y_i)\}_{i=1}^N,
\end{equation}
where $U_i \in \mathbb{R}^{n_u h}$ and $Y_i \in \mathbb{R}^{n_y h}$ denote input and output trajectories of length $h$, respectively, with $h$ representing the time horizon of each experiment.

Since the system is partially measured, we exploit the history of input-output data to represent the unknown input-output relationship. Moreover, since the desired behavior corresponds to periodic motion, we adopt the modeling approach of \cite{banerjee2025data}, which is suited to periodic dynamical systems, and represents the dynamic relationships as a mapping from input-output time series
\begin{equation}\label{eq:msp}
    Y=F(u_0,\dots, u_{t})
\end{equation}
where  $Y$ is defined as $\mathcal{Y}=\begin{bmatrix}
    y_0&y_1&\dots y_t
\end{bmatrix}^T$. However, the main drawback of this approach is the curse of dimensionality when the sampling frequency is high. 

Considering the control objective and the modeling challenges discussed above, we transform the input–output data into the frequency domain using the Fourier transform. The rationale behind this step is that nonlinear systems driven by periodic inputs generate a discrete set of frequency components in steady state, as established in Volterra series theory. For example, a single-frequency input produces harmonic components, whereas a multi-frequency input leads to intermodulation distortion, generating additional frequency components beyond the original inputs and their standard harmonics.

This harmonic structure aligns naturally with the internal model framework, as the steady-state control input can be represented in terms of a finite set of frequency components. Moreover, representing the data in the frequency domain enables dimensionality reduction by concentrating the relevant dynamics into a small number of dominant harmonic coefficients, thereby facilitating structured dataset generation. 

The data set \eqref{eq:data} in the frequency domain can be represented as
\begin{equation}
\label{eq:data_freq}
\mathcal{D}_{\omega} = \{(\hat{U}_i(\omega), \hat{Y}_i(\omega))\}_{i=1}^N,
\end{equation}
where $\hat{U}_i(\omega)$ and $\hat{Y}_i(\omega)$ consist of the real and imaginary parts of the complex values at the frequency bins of interest. Specifically, for an excitation centered at frequency $\omega$, the frequency-domain representations of the input and output are constructed as
\begin{equation}
\hat{U}_i(\omega)=\begin{bmatrix}
    \hat{U}_i(0) \\Re (\hat{U}_i(\omega)) \\ Im(\hat{U}_i(\omega))\\ Re (\hat{U}_i(2\omega)) \\ Im(\hat{U}_i(2\omega)) \\ \vdots
\end{bmatrix}, \;
\hat{Y}_i(\omega)=\begin{bmatrix}
    \hat{Y}_i(0) \\ Re (\hat{Y}_i(\omega)) \\Im(\hat{Y}_i(\omega))\\ Re (\hat{Y}_i(2\omega)) \\ Im(\hat{Y}_i(2\omega)) \\ \vdots
\end{bmatrix}
\end{equation}

Thus, the representation of an unknown system through the data set \eqref{eq:data_freq} boils down to a mapping such that
\begin{equation}
   \hat{Y}(\omega) = F(\hat{U}(\omega))
\end{equation}
where $F$ can be obtained in a certain nonlinear model class $\mathcal{F}$ as
\begin{equation}
  F \in \text{argmin}_{F\in \mathcal{F}} J_r(F)
\end{equation}
where $J_r(F)$ denotes a regression loss defined according to the chosen training procedure. In this work, we utilize neural networks due to their strong approximation capabilities in capturing complex input–output relationships.

\begin{remark}
    This modeling approach is highly practical in settings where systems operate in cyclical or repetitive tasks, such as robotic manipulators, precision manufacturing, and power electronic converters. In such environments, ensuring high-fidelity tracking of periodic trajectories in the presence of unmodeled, bandwidth-limited nonlinearities is critical for performance, making a frequency domain tracking approach both practical and computationally advantageous.
\end{remark}

\subsection{Data-driven Steady-State Generator Design in Frequency Domain}
Our objective is to design a feedforward steady-state generator for an unknown dynamical system such that the tracking error is minimized in a suitable metric for a class of reference signals
\begin{equation}
    y_r(t) = A_0 + \displaystyle\sum_{j=1}^{N_s} A_j \sin(\omega_j t+\phi_j)
\end{equation}
which can also be represented in the form of \eqref{eq: exosystem}. To achieve this objective, we leverage the data-driven modeling approach described in the previous section together with the harmonic representation of the steady-state input given in \eqref{eq: frequency}. This representation shows that the steady-state input lies in the span of the fundamental frequency and its harmonics, with coefficients determined by the amplitudes and phases of the exogenous signals. Consequently, the steady-state generator can be viewed as a mapping from the reference parameters to these harmonic coefficients, and its construction can be formulated as an optimization problem.

%By parameterizing the control input according to the harmonic structure in \eqref{eq: frequency}, the design problem reduces to determining the coefficients of the harmonic components. These coefficients depend on the amplitudes and phases of the exogenous signals. Consequently, by viewing the control input as a mapping from the exogenous signal parameters (amplitudes and phases) to the harmonic coefficients, the construction of the steady-state generator can be formulated as an optimization problem. 

\begin{figure*}
    \centering
    \includegraphics[width=\linewidth]{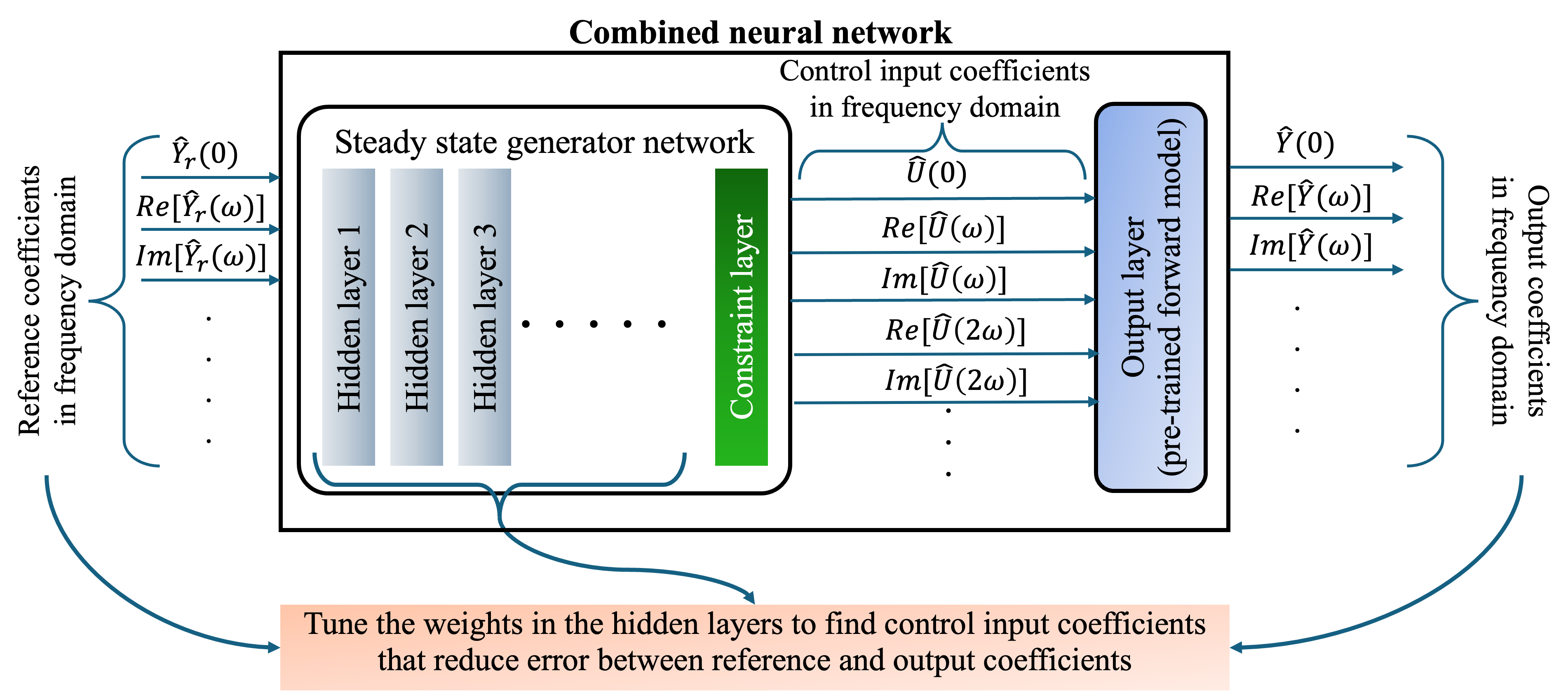}
    \caption{NN-based feedforward controller design}
    \label{fig:NNinversion}
\end{figure*}

To implement this idea, the proposed architecture consists of two cascaded neural networks: a steady-state generator network and a pre-trained forward-model obtained in the model generation stage. The overall architecture is illustrated in Figure \ref{fig:NNinversion}. The pre-trained forward-model network represents the learned frequency-domain mapping of the plant, while the steady-state generator network produces harmonic coefficients based on the reference parameters. The final layer of the generator network incorporates a constraint layer that enforces any imposed constraints. This architecture was inspired by end-to-end constrained optimization learning frameworks~\cite{kotary2021E2ECOL}.

The steady-state generator map takes the reference amplitude and phase information in the frequency domain and produces the corresponding feedforward control input in the frequency domain. By parameterizing this map using a neural network, the proposed design reduces the problem to a learning task, in which the network is trained to approximate the mapping between reference spectral components and the corresponding control input spectral components that minimize the tracking error. This is achieved by minimizing the cost function
\begin{equation}
\label{eq: inversion}
\begin{aligned}
& \underset{\theta}{\min}
& & J_{FF}\big(\hat{Y}_{r}(\omega), \hat{Y}(\omega)\big) \\
& \text{s.t.}
& & \hat{U}(\omega) = G_\theta(\hat{Y}_r(\omega)), \\
&&& \hat{Y}(\omega) = F(\hat{U}(\omega))
\end{aligned}
\end{equation}
where  $G_\theta$ is the steady-state generator map, $\theta$ represents the weights and biases for the map, and the cost function $J_{FF}(\hat{Y}_{r}(\omega),\hat{Y}(\omega))$ is 
\begin{align}\label{eq:cost_function}
    &J_{FF}(\hat{Y}_{r}(\omega),\hat{Y}(\omega)) = \frac{1}{2n_f + 1} ( w_0(\hat{Y}_{r}(0) - \hat{Y}(0))^2 \nonumber \\
    &+ \sum_{i=1}^{n_f} w_i \Bigg[ (\Re (\hat{Y}_r(\omega_i)) - \Re (\hat{Y}(\omega_i)))^2 \nonumber \\
    & + (\Im (\hat{Y}_r(\omega_i)) - \Im (\hat{Y}(\omega_i)))^2 \Bigg] ) 
\end{align}
and $n_f$ is the number of frequencies included in the control design, and $w_j$ are the weights used for focusing the reduction of error for certain frequency components. 

The surrogate model $F$ is trained on inputs from a bounded amplitude and frequency range, and its predictions might degrade outside this range. To keep the generator's output within the validity region of $F$, we introduce a constraint layer at the output of the steady-state generator map. Concretely, this layer can restrict the number of allowable frequency components in the control input and clip the amplitude at each retained component to the range used during surrogate training. Physical prior knowledge can also inform the constraint design, for example, by imposing actuator limits or excluding frequencies outside the achievable bandwidth.

The proposed data-driven approach admits a clean interpretation in classical IMP terms: the trained steady-state generator map $G_\theta$ is a data-driven realization of the frequency domain representation of the ideal feedforward control input $c(v)$ that the regulator equations \eqref{eq: reg_eqn1}-\eqref{eq: reg_eqn2} would produce in the traditional model-based setting. The harmonic parameterization adopted here is motivated by the equivalence established in \cite{huang2001remarks}: when $c(v)$ admits a polynomial representation in the exosystem states, the steady-state input is supported on a finite set of harmonics of the fundamental frequencies. Our framework exploits this harmonic structure as the natural basis for the steady-state generator, while not requiring $c(v)$ itself to be polynomial. As demonstrated in Section 4, non-polynomial nonlinearities also produce harmonic content that is concentrated, up to truncation, in a finite set of bins; the proposed method recovers the corresponding coefficients implicitly through the learned plant model $F$, without solving the regulator equations.

%In addition to the hidden layers in the steady-state generator map, a constraint layer is also included at the end to keep the control inputs produced in the reliable regions of the forward model, reducing the variability in tracking performance. This includes limiting the number of allowable frequency components, amplitude range for each frequency components. 

%\begin{equation}
%\label{eq: inversion}
%\begin{aligned}
%& \underset{\hat{U}(\omega)}{\min}
%& & J_{FF}(\hat{Y}_{r}(\omega),\hat{Y}(\omega)) \\
%& \text{s.t.}
%& & \hat{U}(\omega) \in \mathcal{U}
%\end{aligned}
%\end{equation}
\section{Numerical Results}
To demonstrate the efficacy of the proposed approach, we consider two numerical case studies: a mechanical load with nonlinear friction, driven by a DC motor, borrowed from \cite{yao2015output}, and an electrohydraulic (EH) actuator system.
The first example serves as a benchmark, as analytical solutions to the corresponding regulator equations are readily obtainable, enabling a direct comparison between the data-driven solution and the exact analytical result. The EH actuator system is considered as a second example to assess the generalizability of the approach.
\subsection{Mechanical load with nonlinear friction}
Consider the following second-order mechanical system
\begin{align}
    \label{eq: DC1}
    \dot{x}_1 &= x_2\\
    \dot{x}_2 &= a_1u - a_1 F_{fric}(x_2) - a_2 x_1\\
    y &= x_1
    \label{eq: DC2}
\end{align}
where $x_1$ and $x_2$ denote position and velocity, respectively, and $u$ represents the control input. The term $F_{fric}(x_2)$ models smooth nonlinear friction and is given by
\begin{align}
\label{eq: friction}
    F_{fric}(x_2) &= \alpha_1\tanh (\beta_1x_2) \nonumber \\
     & + \alpha_2 \big[\tanh(\beta_2 x_2)-\tanh(\beta_3 x_2)\big] + \alpha_3 x_2
\end{align}
where the term $\alpha_1\tanh (\beta_1x_2)$ represents the Coulomb friction, the second term captures the Stribeck effect and $\alpha_3 x_2$ is the viscous friction. The parameters are given in Table \ref{tab: Mechparam}.

The control objective is to ensure that the output $y$ tracks a periodic reference signal $y_r(t)=A\sin(\omega t)$, generated by the exosystem
\begin{align*}
    \begin{bmatrix}
        \dot{v}_1\\
        \dot{v}_2
    \end{bmatrix}
    &=
    \begin{bmatrix}
        0 & \omega\\
        -\omega & 0
    \end{bmatrix}
    \begin{bmatrix}
        v_1\\
        v_2
    \end{bmatrix}, \quad
    v(0)=
    \begin{bmatrix}
        0 \\ A
    \end{bmatrix}, \\
    y_r &= v_1.
\end{align*}
To obtain the ideal feedforward input, we substitute the plant model \eqref{eq: DC1}--\eqref{eq: DC2} into the regulator equations \eqref{eq: reg_eqn1}-\eqref{eq: reg_eqn2}, and solve analytically for the steady-state input $u = c(v)$
\begin{equation}
\label{eq: analytic}
    c(v) =  \frac{a_2-\omega^2}{a_1}v_1 + F_{fric}(\omega v_2) 
\end{equation}

The analytical solution $c(v)$ exhibits the harmonic structure as anticipated by Section 3.2. The linear term in \eqref{eq: analytic} contributes at the fundamental frequency, and the nonlinear term, generated by $F_{fric}(\omega v_2) $, introduces additional harmonics induced by the friction nonlinearity. The proposed data-driven generator is therefore expected to recover the dominant components of this structure without prior knowledge of the plant dynamics.

The proposed approach requires a surrogate model, as described in the previous section. 
Thus, training data are generated by simulating the model \eqref{eq: DC1} - \eqref{eq: DC2} under input signals composed of three sinusoidal components selected from 
 $\{0, 1, 2, 3, 4, 5\}$ Hz, where 0 Hz denotes the DC term. The choice to parameterize each input by three sinusoidal components reflects a key property of the proposed framework: the harmonic content required for accurate tracking is not assumed to be known a priori. The training data, therefore, cover all possible combinations of active components, and the steady-state generator implicitly learns which combination is appropriate for a given reference. For each combination, amplitudes are sampled from $(0,\; 0.10)$ for the DC term and $(0,\; 0.35)$ for each active sinusoidal component, with phases sampled from $(0,\; 2\pi)$.  The amplitude–phase space is covered using Latin Hypercube Sampling (LHS), yielding a total of 1.28 million data points. Then, this data set is used to train a neural network surrogate model with five hidden layers consisting of 256 neurons in each layer and the ReLU activation function. The input-output dimension of the neural network is 11, comprising the DC component and the real and imaginary parts at 1, 2, 3, 4, and 5 Hz. Training is performed in PyTorch using the Adam optimizer with 1000 epochs at $10^{-3}$ learning rate and $10^{-4}$ weight decay. The trained model achieves a mean-square error (MSE) of $1.22 \times 10^{-4}$ and $2.13 \times 10^{-4}$ $deg^2$ on the train and test data set.

The steady-state generator map is then constructed by incorporating the surrogate model into the training procedure. The training data for this stage consists of reference signals at 1 Hz with amplitudes and phases sampled from the ranges $(0.1,\; 0.22)\ rad$ and $(0,\;2\pi)\ deg$. The training data has a total of 40,000 reference signal combinations. A multi-layer neural network is employed to map the frequency-domain representation of the reference signal to the harmonic components required for feedforward tracking. This network is parameterized by four hidden layers with sizes $[64, 128, 128, 64]$ and the ReLU activation function, with an input-output dimension of 11. The training is performed for 1000 epochs with $10^{-4}$ learning rate and the resulting MSE is $2.18\times 10^{-6}$ $deg^2$ for the test data.

\begin{figure}
     \centering
     % --- First Subfigure ---
     \begin{subfigure}[b]{0.45\textwidth}
         \centering
         \includegraphics[width=\textwidth]{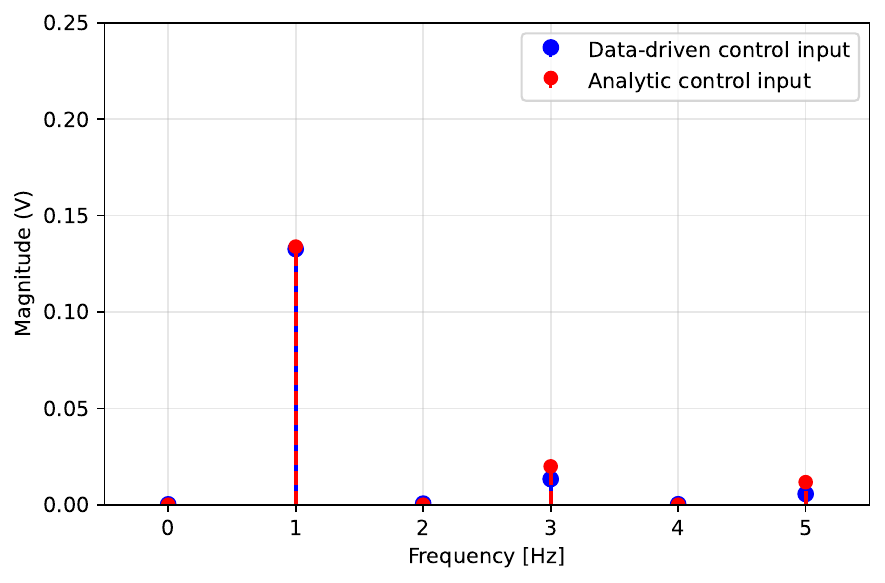}
         \caption{Control input for reference signal $y_r^1(t)$}
         \label{fig: comp_sub1}
     \end{subfigure}
     \hfill
     % --- Second Subfigure ---
     \begin{subfigure}[b]{0.45\textwidth}
         \centering
         \includegraphics[width=\textwidth]{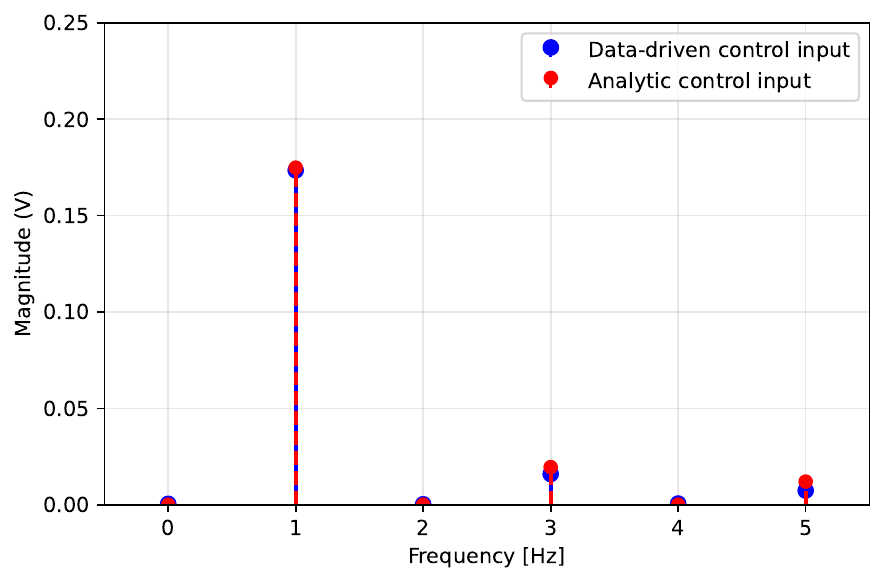}
         \caption{Control input for reference signal $y_r^2(t)$}
         \label{fig: comp_sub2}
     \end{subfigure}
     \hfill
     \caption{Comparison of the analytic (red) and the data-driven feedforward input (blue) for the mechanical load with nonlinear friction in frequency domain.}
     \label{fig: verification}
\end{figure}

To validate the proposed approach, we first present a frequency domain comparison between the analytic solution $c(v)$ and the data-driven solution. Since the nonlinear friction term is non-polynomial, the corresponding steady-state feedforward input generally contains infinitely many harmonic components. For this comparison, we retain frequency components up to 5 Hz to evaluate whether the proposed method can recover the dominant harmonic structure within the selected bandwidth. Figure \ref{fig: verification} illustrates this comparison for two reference signals, 
\begin{align*}
    y_r^1(t) &= 7\sin(2\pi t + 0.25\pi)\ deg\\
    y_r^2(t) &= 10\sin(2\pi t)\ deg.
\end{align*}
It can be observed from the analytical solution that the feedforward control input requires a harmonic component at 3 and 5 Hz in addition to the fundamental frequency of the reference trajectory at 1 Hz. On the other hand, the data-driven solution accurately captures the harmonic structure as a nonlinear relationship between them required to achieve tracking. 

% \begin{figure}
%     \centering
%     \includegraphics[width=0.80\linewidth]{images/NNInversionController_v3.png}
%     \caption{NN-based inversion controller design}
%     \label{fig:NNinversion}
% \end{figure}

\begin{figure}
     \centering
     % --- First Subfigure ---
     \begin{subfigure}[b]{0.46\textwidth}
         \centering
         \includegraphics[width=\textwidth]{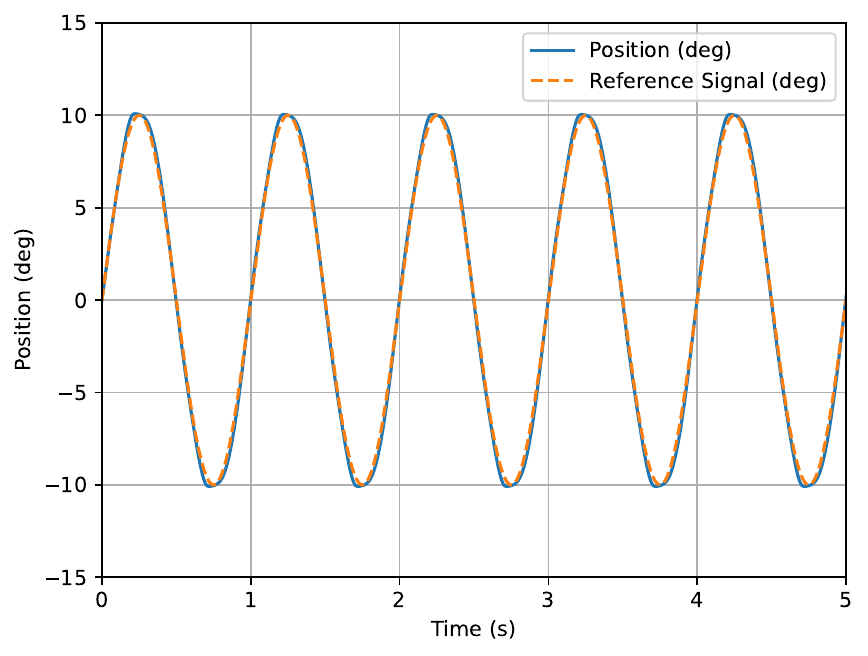}
         \caption{Reference and system response using the data-driven control input in time domain }
         \label{fig: sub1}
     \end{subfigure}
     \hfill
     % --- Second Subfigure ---
     \begin{subfigure}[b]{0.44\textwidth}
         \centering
         \includegraphics[width=\textwidth]{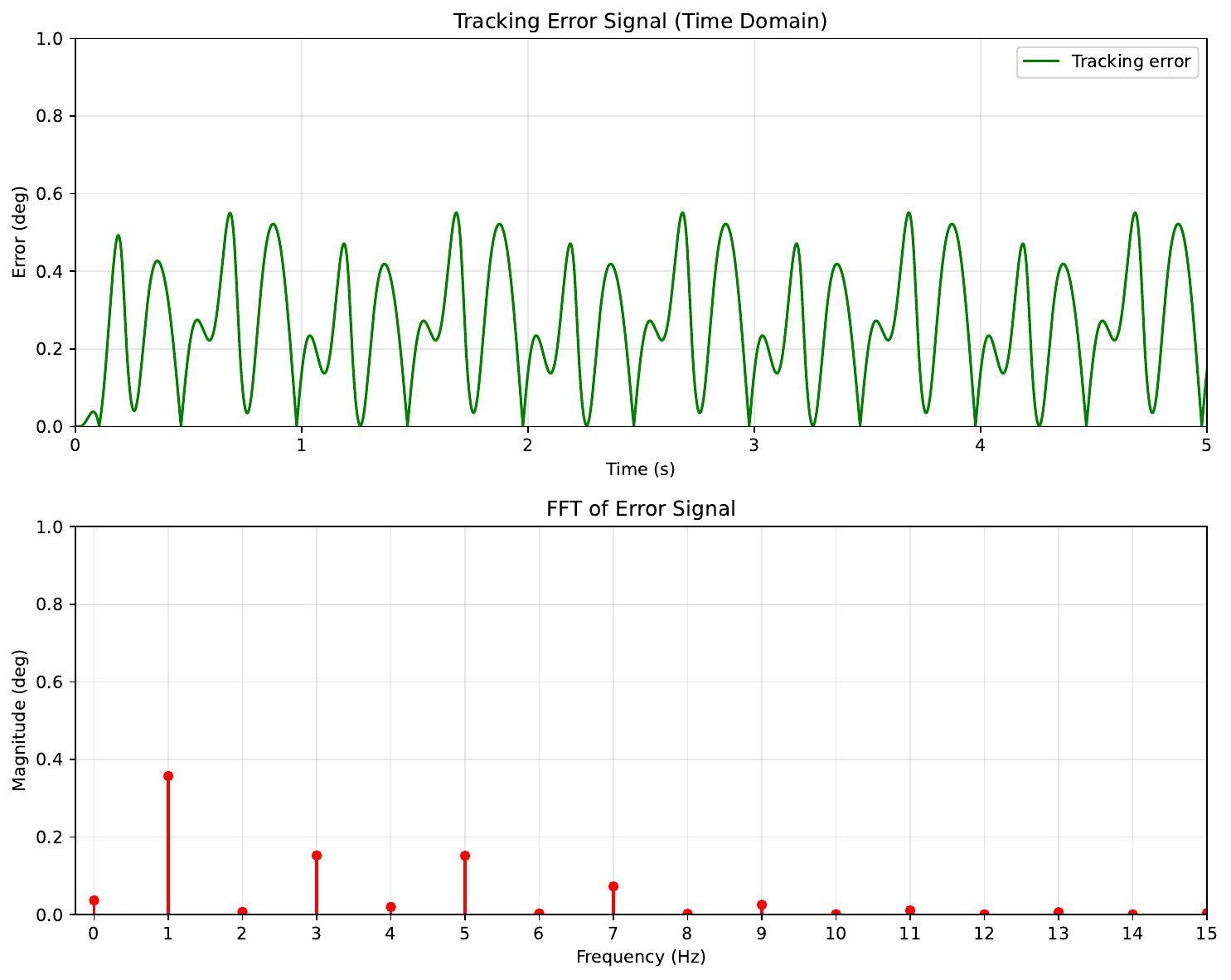}
         \caption{Time and frequency domain tracking error}
         \label{fig: sub2}
     \end{subfigure}
     \hfill

     \caption{Tracking performance in time and frequency domain for the mechanical load with nonlinear friction for $y_r^1(t)$}
     \label{fig: overall_tracking_results_mech}
\end{figure}

We further present both time domain and frequency domain tracking error results obtained from the implementation of the data-driven feedforward control input to the physics-based model  \eqref{eq: DC1}-\eqref{eq: DC2} in Figure \ref{fig: sub1}-\ref{fig: sub2}. For the 10 deg reference trajectory, the truncated analytical feedforward input yields a maximum tracking error of 0.36 deg, while the proposed data-driven feedforward input yields 0.55 deg. Although the data-driven solution introduces a modest increase in peak error, the error remains $5.5\%$ of the reference amplitude and is close to the truncated analytical benchmark. This demonstrates that the proposed method can recover the dominant harmonic structure required for tracking without explicitly relying on the physics-based model.

%It is worth noting that the nonlinear friction term (involving $\tanh(\cdot)$) is non-polynomial and, in principle, generates an infinite number of odd harmonics. However, due to the limited bandwidth of the system, higher-order harmonics have negligible influence on the system response and can be truncated during the control design process.

\begin{table}
    \centering
    \begin{tabular}{|c|c|c|}
    \hline
    Parameters & Value & Units \\ 
    \hline
         $a_1$&  400 & $rad/(V\cdot s^{-2})$ \\
         $a_2$&  355 & $s^{-2}$ \\
         $\alpha_1$&  0.04 & $V$\\
         $\alpha_2$&  0.01 & $V$\\
         $\alpha_3$& 0.05 & $V\cdot s/rad$\\
         $\beta_1$& 15 & $s/rad$\\
         $\beta_2$& 15 & $s/rad$\\
         $\beta_3$& 1.5 & $s/rad$\\
         \hline
    \end{tabular}
    \caption{Parameters used in the mechanical system and the friction term}
    \label{tab: Mechparam}
\end{table}

\subsection{Electrohydraulic actuator system}

\begin{figure}
    \centering
    \includegraphics[width=\linewidth]{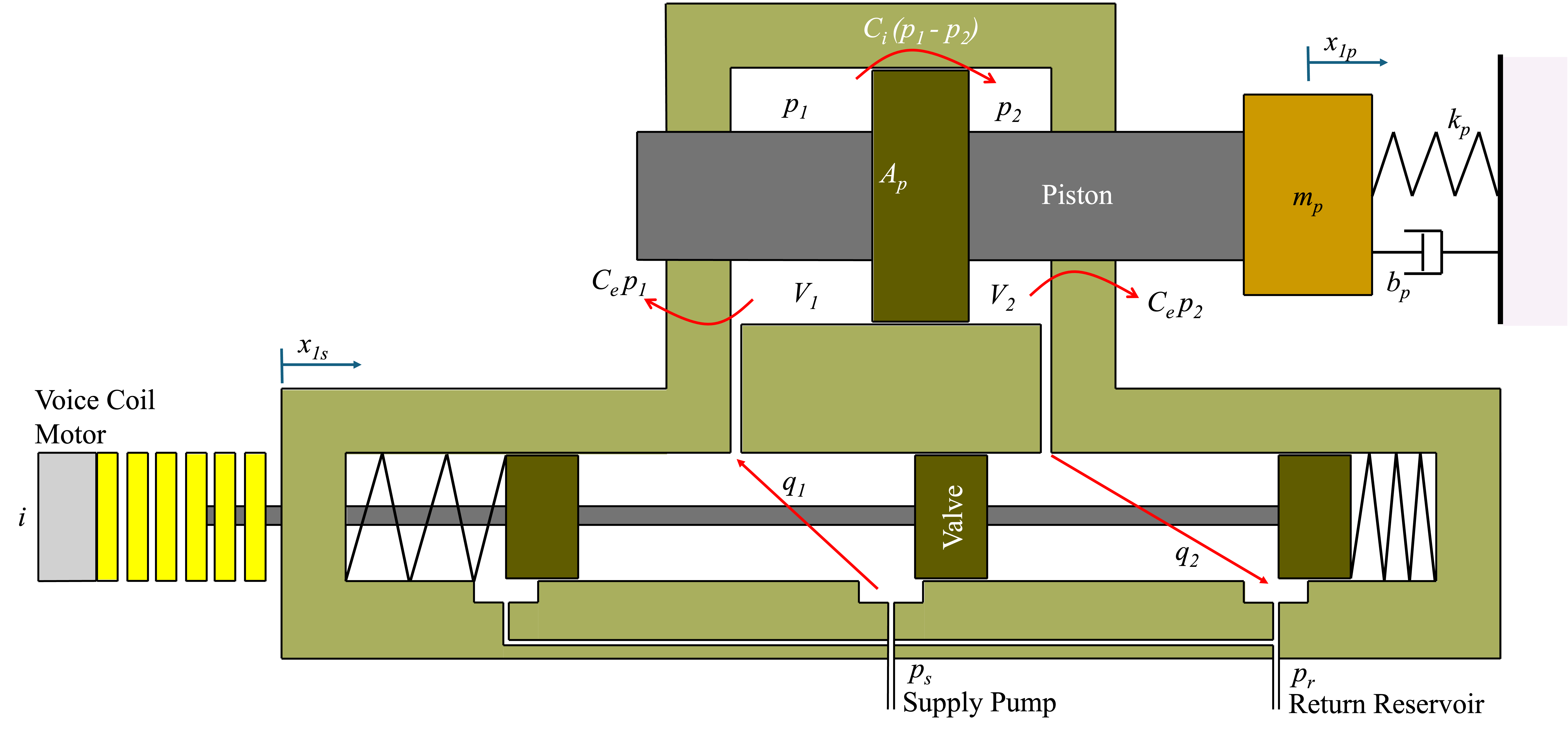}
    \caption{Electrohydrualic actuator schematic}
    \label{fig:EHAsystem}
\end{figure}

EH actuator systems are widely used in applications that require precise motion control under high-load conditions, including aerospace, automotive, and industrial robotic systems. A major challenge in tracking control design of EH actuators arises from the nonlinear flow-pressure relationships imposed by the valve orifice equations. The governing equations of the EH actuator system are given by \cite{Yoon2019-pa}:
\begin{align}
\label{eq: EH1}
\dot{x}_{1p}(t)&=x_{2p}(t),\\
\dot{x}_{2p}(t)&=\frac{1}{m_p}\big(-b_p x_{2p}(t) \nonumber\\
            & - k_p x_{1p}(t) + A_p(p_1(t)-p_2(t))\big),\\
\dot{p}_1(t) &= \frac{\beta}{V_1(t)}
\big(q_1(t)-A_p x_{2p}(t) - \nonumber\\
& C_i(p_1(t)-p_2(t)) - C_e p_1(t) \big),\\
\dot{p}_2(t) &= \frac{\beta}{V_2(t)}
\big(-q_2(t)+A_p x_{2p}(t) + \nonumber\\
& C_i(p_1(t)-p_2(t)) - C_e p_2(t) \big),
\end{align}
where \(x_{1p}(t)\), \(x_{2p}(t)\), \(p_1(t)\), and \(p_2(t)\) denote the piston position, piston velocity, left chamber pressure, and right chamber pressure, respectively. The left and right chamber volumes are \(V_1(t) = V_0+A_p x_{1p}(t)\) and \(V_2(t) = V_0-A_p x_{1p}(t)\) respectively. The flow rates into the left and right chambers are denoted by \(q_1(t)\) and \(q_2(t)\), respectively, and are given by
\begin{align}
q_1(t) &=
\begin{cases}
C_o x_{1s}(t)\sqrt{2(P_s-p_1(t))}, & x_{1s}(t)\geq 0,\\
C_o x_{1s}(t)\sqrt{2(p_1(t)-P_r)}, & x_{1s}(t)<0,
\end{cases}\\
q_2(t) &=
\begin{cases}
C_o x_{1s}(t)\sqrt{2(p_2(t)-P_r)}, & x_{1s}(t)\geq 0,\\
C_o x_{1s}(t)\sqrt{2(P_s-p_2(t))}, & x_{1s}(t)<0,
\end{cases}
\end{align}
where \(C_o=C_d W/\sqrt{\rho}\). The spool position \(x_s(t)\) is governed by the valve dynamics
\begin{align}
    \dot{x}_{1s}(t) &= x_{2s}(t),\\
    \label{eq: spool2}
    \dot{x}_{2s}(t) &= -2\zeta \omega_n x_{2s}(t) - \omega_n^2 x_{1s}(t) + k_b\omega_n^2 i(t),
\end{align}
where \(i(t)\) is the solenoid current, which is the control input. The available measurment is assumed to be the piston position ($x_{1p}(t)$). The system parameters are given in Table \ref{tab:EHAparam}. A schematic representation of the EH actuator system is shown in Figure~\ref{fig:EHAsystem}.

\begin{table}
    \centering
    \begin{tabular}{|p{1.4cm}|p{2.2cm}|p{1.3cm}|p{1.4cm}|}
    \hline
    Parameters & Definition &  Value & Units\\ 
    \hline
         $m_p$& Piston mass & 10 & $kg$\\
         $b_p$&  Damping coefficient &500& $Ns/m$\\
         $k_p$&  Spring stiffness &  2e5 & $N/m$\\
         $A_p$& Piston cross-sectional area & 1.5e-3 &$m^2$\\
         $\beta$& Bulk modulus &7e8 & $Pa$\\
         $V_0$& Volume at the neutral position ($x_{1p}=0$ )&2.945e-4 &$m^3$\\
         $C_i$& Internal leakage coefficient &2e-12& $m^3/(s.Pa)$\\
         $C_e$& External leakage coefficient &1e-12 &$m^3/(s.Pa)$\\
         $C_d$& Discharge coefficient &0.61& -\\
         $\zeta$& Damping factor of the spool valve &0.707& - \\
         $\omega_n$& Natural frequency of the spool valve &125.66& $rad/s$\\
         $k_b$& Voice coil motor scaling factor &0.004 & $m/A$\\
         $W$ & Area gradient & 0.0785 & $m$ \\
         $\rho$ & Oil density & 870 & $kg/m^3$ \\
         $P_s$& Supply pressure &10e6 & $Pa$\\
         $P_r$& Return pressure &1e6 & $Pa$\\
         \hline
    \end{tabular}
    \caption{Parameters used in the Electro Hydraulic actuator (EHA) system}
    \label{tab:EHAparam}
\end{table}

As described in Section 3.2, the first step is to obtain a surrogate model that represents the input–output relationship in the frequency domain. To this end, data are collected through numerical simulations of the physics-based model \eqref{eq: EH1}-\eqref{eq: spool2}. The input is parameterized using combinations of up to four frequency components selected from the set $\{0, 5, 10, 15, 20, 25, 30\}$ Hz with varying amplitudes $(0,\; 0.1)$ A for the DC term, and $(0,\; 0.5)$ A for the remaining components. The phase range is chosen as $(0,\; 2\pi)$. Sampling over the prescribed amplitude–phase ranges is performed using LHS. Accordingly, a total of 4.54 million control input signals are generated.

The data set obtained through numerical simulations was utilized to train a neural network, which serves as a surrogate data-driven model of the EH actuator system. The network was implemented with seven hidden layers of sizes $[ 50, 75, 100, 100, 100, 75, 50]$ with a $\tanh$ activation function applied after each hidden layer. 
Training is performed in PyTorch using the Adam optimizer for 200 epochs, with a learning rate $10^{-3}$ for the first 100 epochs and $5\times10^{-4}$ for the remaining 100. The final training and validation MSE values were 0.047 mm$^2$ and 0.051 mm$^2$, respectively.

The control objective is for the piston position to track reference signals consisting of a 5 Hz fundamental with 50 mm amplitude and its higher harmonics. Following the training procedure described in Section 3.2, we incorporate the surrogate model into the cascaded training architecture and obtain the steady-state generator map $G_{\theta}$ for the EH actuator system. Specifically, the generator takes a 13-dimensional vector representing the reference signal in the frequency domain - comprising the DC component and the real and imaginary parts at 5, 10, 15, 20, 25, and 30 Hz — and produces a 13-dimensional vector.

The generator map is a multi-layer perceptron with five hidden layers of sizes $[25,50,75,50,25]$, with input and output dimensions of 13. The activation function $\tanh$ is applied after each hidden layer. Since the control objective is to track reference signals consisting of a 5 Hz fundamental and its higher harmonics, with the fundamental amplitude up to 50 mm, the training references are constructed to span this class. Each reference signal contains the 5 Hz fundamental and at most two additional harmonics drawn from the set $\{0, 10, 15, 20, 25, 30\}$ Hz; harmonic amplitudes decay faster than the $1/n$ rate of a square wave. Phases are sampled from the range $(0,\ 2\pi)$. Approximately 5 million training references are generated using LHS. Training is performed using the Adam optimizer for 300 epochs with a learning rate of 0.001. In addition, the weight $w_i$ in the cost function \eqref{eq:cost_function} for the 5 Hz frequency component is selected as 1.5 to prioritize reducing the tracking error at the fundamental frequency. The final MSE for training and validation datasets was 0.227 $mm^2$ and 0.226 $mm^2$ with the data-driven forward model.

A constraint layer is applied at the output of the generator at inference time. This layer enforces two constraints: the control input contains at most four harmonic components, including all frequencies present in the reference; and the magnitudes of these components are clipped to the amplitude ranges used during surrogate training. These constraints keep the control input within the surrogate's validity envelope while allowing harmonics beyond those in the reference to compensate for nonlinearities induced by the plant.

\begin{figure}
     \centering
     % --- First Subfigure ---
     \begin{subfigure}[b]{0.5\textwidth}
         \centering
         \includegraphics[width=\textwidth]{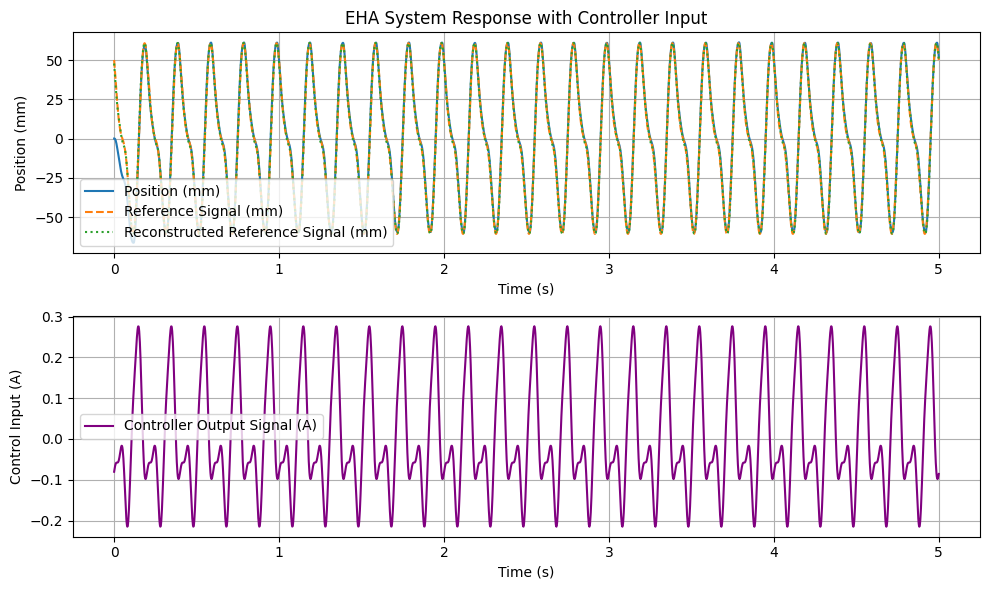}
         \caption{Time domain tracking}
         \label{fig:2harmtrackTimeDomain}
     \end{subfigure}
     \hfill
     % --- Second Subfigure ---
     \begin{subfigure}[b]{0.5\textwidth}
         \centering
         \includegraphics[width=\textwidth]{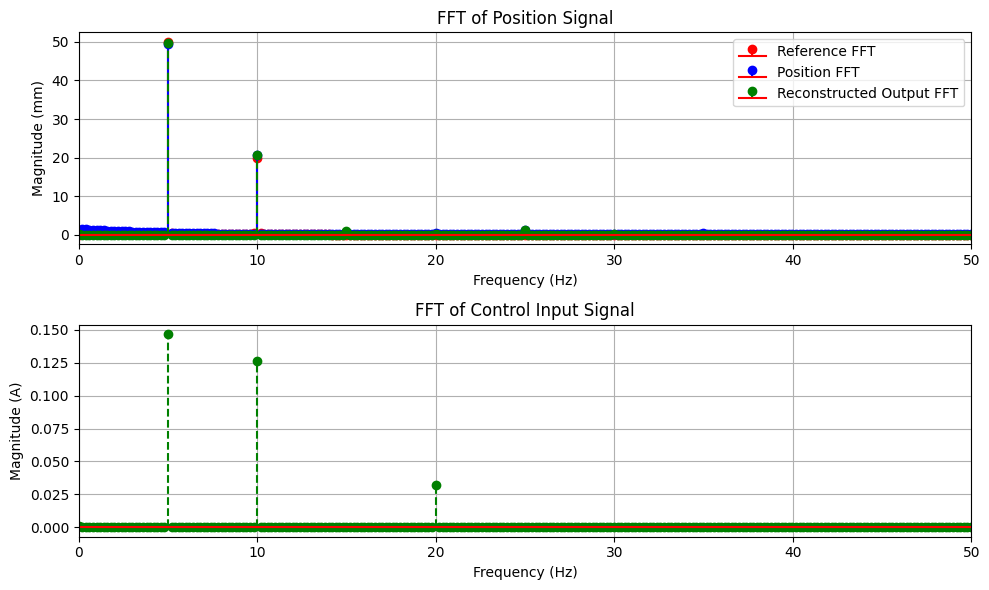}
         \caption{Frequency domain tracking}
         \label{fig:2harmtrackFreqDomain}
     \end{subfigure}
     \hfill
     % --- Third Subfigure ---
     \begin{subfigure}[b]{0.5\textwidth}
         \centering
         \includegraphics[width=\textwidth]{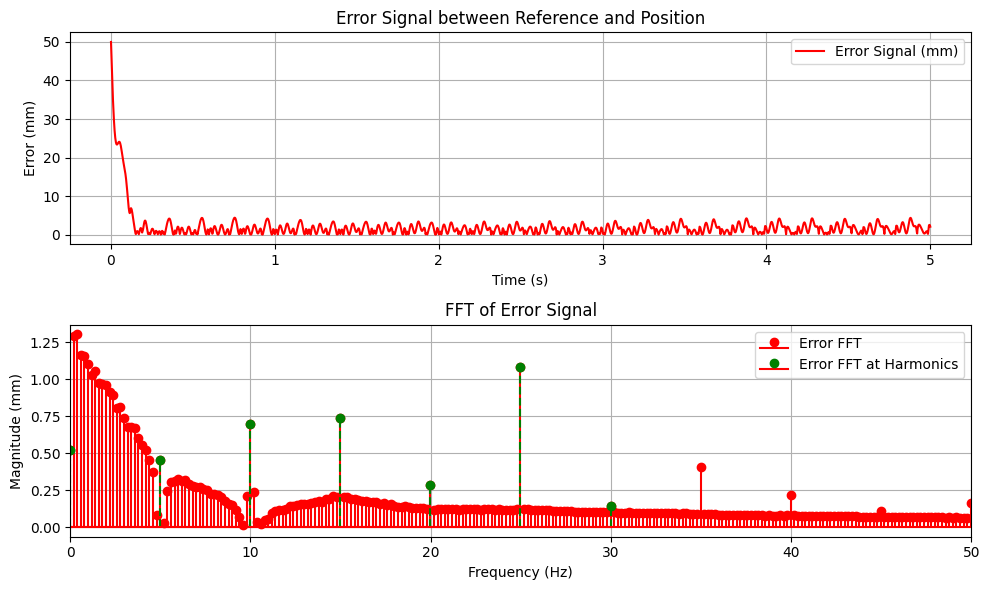}
         \caption{Tracking error}
         \label{fig:2harmError}
     \end{subfigure}

     \caption{Tracking performance and control signals for a reference signal with 2 harmonics for the EHA system.}
     \label{fig:overall_tracking_results_2harmonics}
\end{figure}

\begin{figure}
     \centering
     % --- First Subfigure ---
     \begin{subfigure}[b]{0.5\textwidth}
         \centering
         \includegraphics[width=\textwidth]{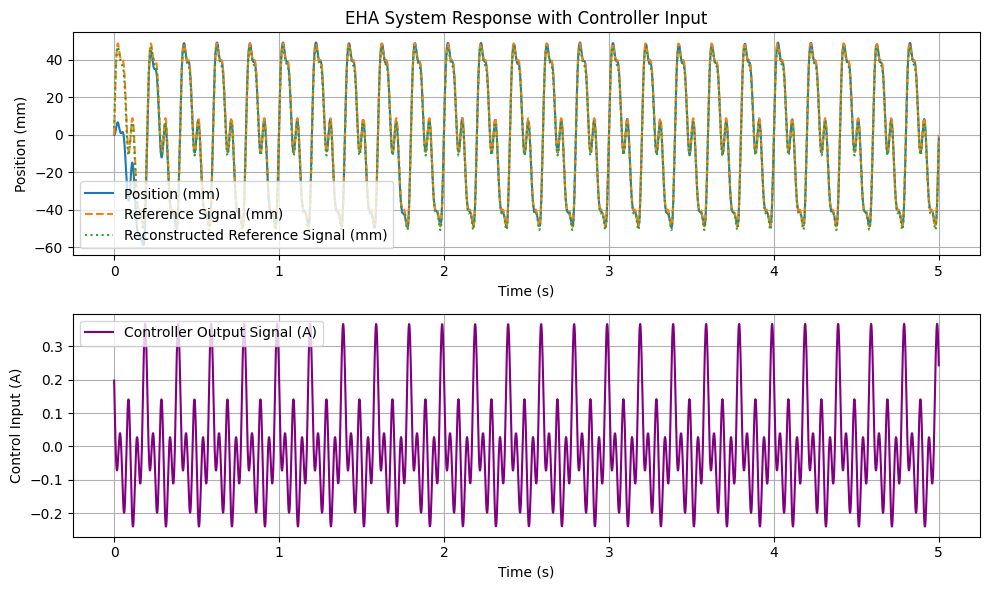}
         \caption{Time domain tracking}
         \label{fig:3harmtrackTimeDomain}
     \end{subfigure}
     \hfill
     % --- Second Subfigure ---
     \begin{subfigure}[b]{0.5\textwidth}
         \centering
         \includegraphics[width=\textwidth]{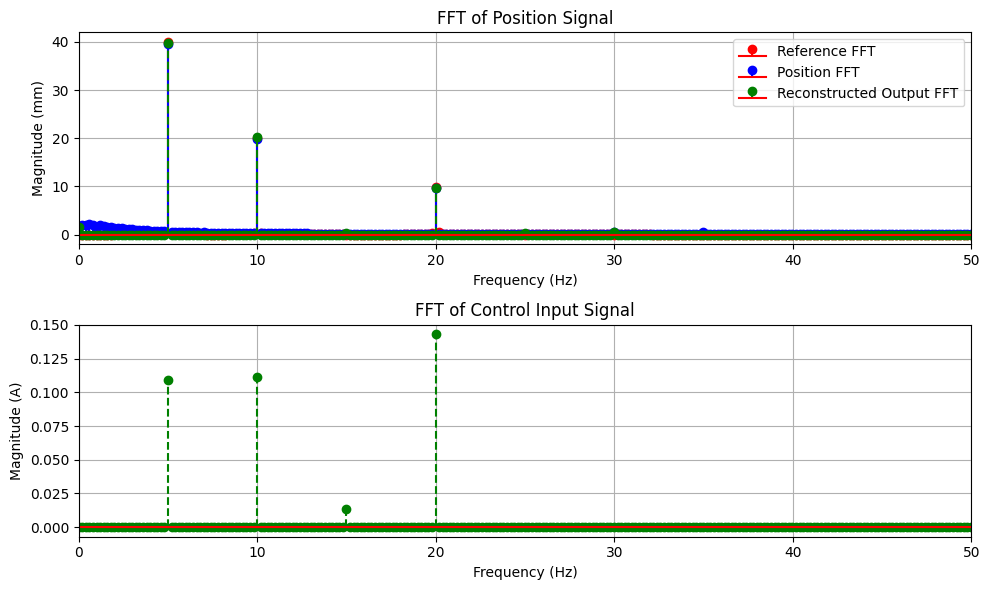}
         \caption{Frequency domain tracking}
         \label{fig:3harmtrackFreqDomain}
     \end{subfigure}
     \hfill
     % --- Third Subfigure ---
     \begin{subfigure}[b]{0.5\textwidth}
         \centering
         \includegraphics[width=\textwidth]{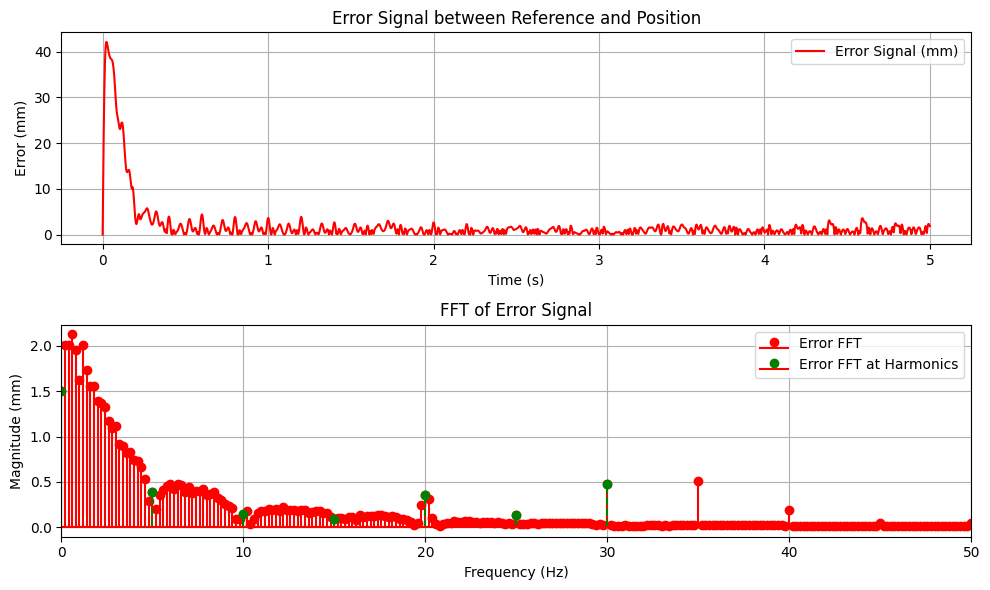}
         \caption{Tracking error}
         \label{fig:3harmError}
     \end{subfigure}

     \caption{Tracking performance and control signals for a reference signal with 3 harmonics for the EHA system.}
     \label{fig:overall_tracking_results_3harm}
\end{figure}

\begin{figure}
     \centering
     % --- First Subfigure ---
     \begin{subfigure}[b]{0.5\textwidth}
         \centering
         \includegraphics[width=\textwidth]{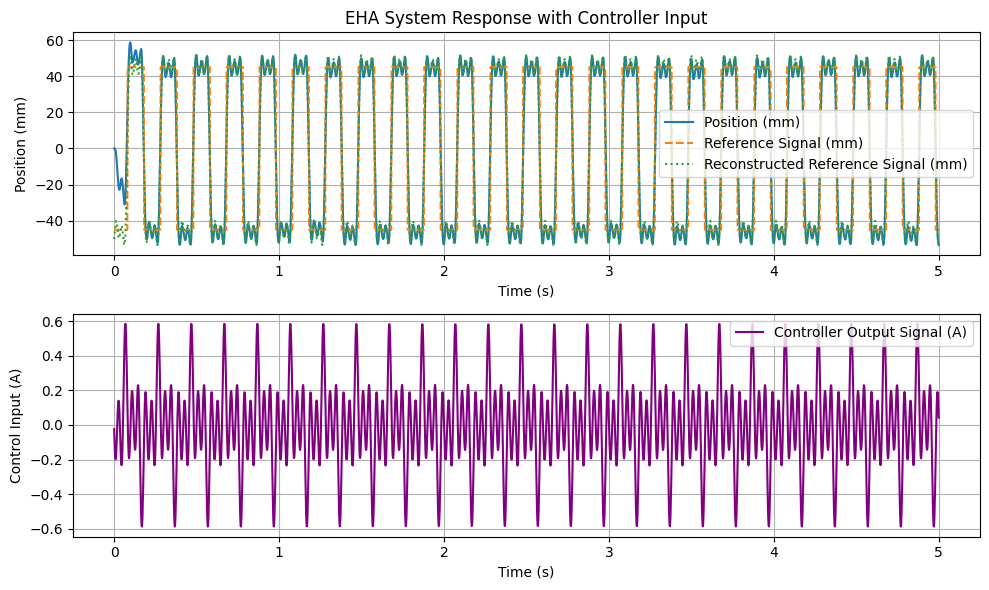}
         \caption{Time domain tracking}
         \label{fig:SquaretrackTimeDomain}
     \end{subfigure}
     \hfill
     % --- Second Subfigure ---
     \begin{subfigure}[b]{0.5\textwidth}
         \centering
         \includegraphics[width=\textwidth]{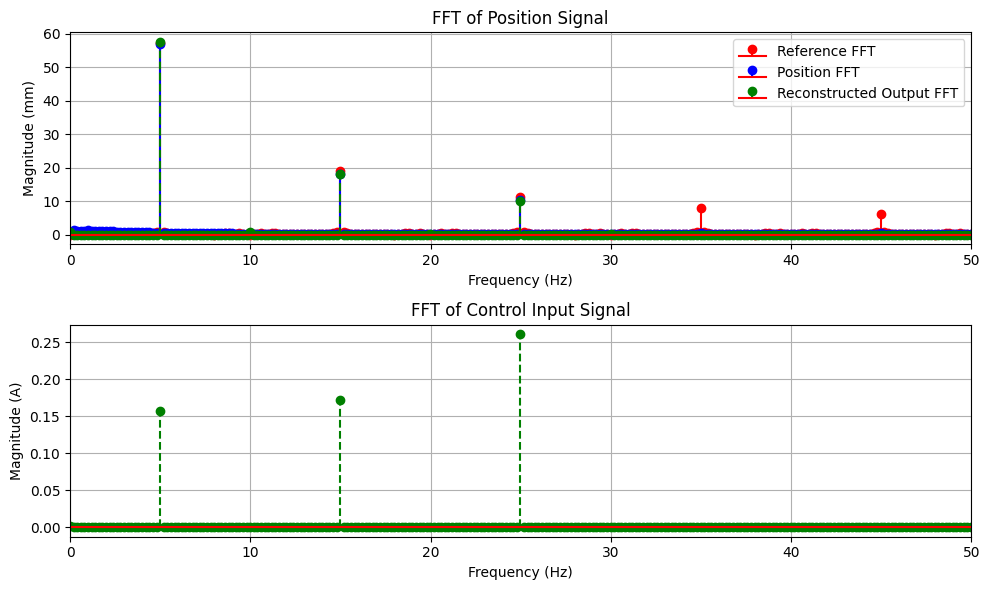}
         \caption{Frequency domain tracking}
         \label{fig:SquaretrackFreqDomain}
     \end{subfigure}
     \hfill
     % --- Third Subfigure ---
     \begin{subfigure}[b]{0.5\textwidth}
         \centering
         \includegraphics[width=\textwidth]{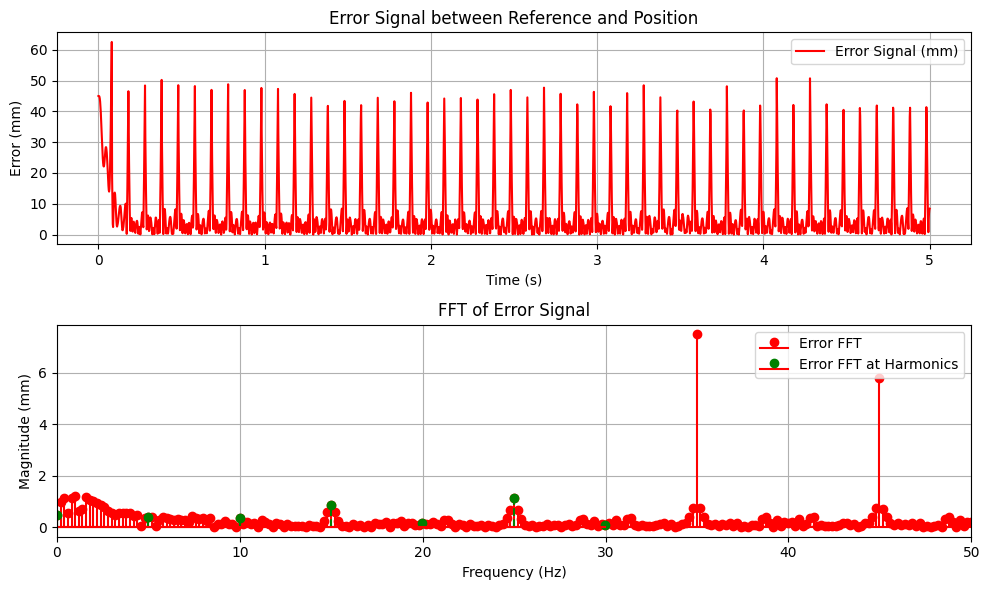}
         \caption{Tracking error}
         \label{fig:SquareError}
     \end{subfigure}

     \caption{Tracking performance and control signals for a square wave reference signal for the EHA system.}
     \label{fig:overall_tracking_results_square}
\end{figure}

The tracking performance of the data-driven design is shown for the following reference signals 
\begin{align*}
    y_r^1(t) &= 50\sin(10\pi t + 0.5\pi) + 20\sin(20\pi t + \pi)\ mm\\
    y_r^2(t) &= 40\sin(10\pi t) + 20\sin(20\pi t) + 10\sin(40\pi t)\ mm \\
    y_r^3(t) &= 45\text{ sgn}(\sin(10\pi t + 1.2\pi))\ mm
\end{align*}
and Figures~\ref{fig:overall_tracking_results_2harmonics}, \ref{fig:overall_tracking_results_3harm}, and \ref{fig:overall_tracking_results_square} show the tracking performance and errors, control signals for reference signals $y_r^1(t)$, $y_r^2(t)$, and $y_r^3(t)$. For each case, the generated feedforward inputs are validated on the physics-based model \eqref{eq: EH1} - \eqref{eq: spool2}. Figures \ref{fig:2harmtrackTimeDomain}, \ref{fig:3harmtrackTimeDomain}, and \ref{fig:SquaretrackTimeDomain} demonstrate the tracking performance (top) and the control inputs (bottom) generated from the controller for each reference signal in time domain. Figures \ref{fig:2harmtrackFreqDomain}, \ref{fig:3harmtrackFreqDomain}, and \ref{fig:SquaretrackFreqDomain} demonstrate the tracking performance (top) and the control inputs (bottom) generated from the controller for each reference signal in frequency domain. 

For each of the cases, the reference signal is first given as an input to the controller to obtain the control input. This control input is then provided as an input to the simulation model as well as the data-driven model built in the earlier section. Control inputs generated from the developed controller 

In the time (Figures~\ref{fig:2harmtrackTimeDomain}, \ref{fig:3harmtrackTimeDomain}, \ref{fig:SquaretrackTimeDomain}) and frequency (Figures~\ref{fig:2harmtrackFreqDomain}, \ref{fig:3harmtrackFreqDomain}, \ref{fig:SquaretrackFreqDomain}) domain tracking plots the output obtained from the simulation model is indicated in blue and the output obtained from the DD-model is shown in green. The closeness of output from the simulation model and the data-driven model demonstrates the ability of the data-driven modeling methodology to accurately capture the underlying dynamics of the system.

The tracking error with respect to the output from the simulation model from the reference signal in time (top) and frequency (bottom) domains are shown in Figure~\ref{fig:2harmError}, \ref{fig:3harmError} and \ref{fig:SquareError}. In all the cases shown, the underlying nonlinearity of the model can be observed where harmonics other than the one provided in the control signal are observed. In Figure~\ref{fig:2harmError}, it can be seen that with four allowable frequency components, the controller is able to reduce the error to be less than 1 mm at those respective components and a maximum time-domain tracking error of 4.3 mm after reaching steady state ($t > 2$ secs). Larger errors are observed at higher frequencies, for which more frequency components would be required. In Figure~\ref{fig:3harmError}, the controller reduces the error using four frequency components in the control signal and the error at these four components is less than 0.5 mm and a maximum time domain tracking error of 3.5 mm after reaching steady state ($t > 2$ secs). To reduce the error further, more frequency components are required. In the case of square wave tracking shown in Figure~\ref{fig:overall_tracking_results_square}, the error of all the frequency components below 30 Hz is lower than 1.5 mm, and the frequency range of the controller needs to be expanded to further reduce the tracking error. The MSE by simulating the control inputs from the final controller on the simulation model in time domain after reaching steady state ($t>2$ secs) for training and validation datasets are 3.19 and 3.18 $mm^2$. The MSE in frequency domain at the $\{0, 5, 10, 15, 20, 25, 30\}$ Hz for training and validation datasets are 0.39 $mm^2$ respectively.

% \begin{figure}[htbp]
%      \centering
%      % --- First Subfigure ---
%      \begin{subfigure}[b]{0.32\textwidth}
%          \centering
%          \includegraphics[width=\textwidth]{images/3harmonic_4000_1300_0800_timeDomainTracking.png}
%          \caption{Time domain tracking}
%          \label{fig:3harmtrackTimeDomain}
%      \end{subfigure}
%      \hfill
%      % --- Second Subfigure ---
%      \begin{subfigure}[b]{0.32\textwidth}
%          \centering
%          \includegraphics[width=\textwidth]{images/3harmonic_4000_1300_0800_freqDomainTracking.png}
%          \caption{Frequency domain tracking}
%          \label{fig:3harmtrackFreqDomain}
%      \end{subfigure}
%      \hfill
%      % --- Third Subfigure ---
%      \begin{subfigure}[b]{0.32\textwidth}
%          \centering
%          \includegraphics[width=\textwidth]{images/3harmonic_4000_1300_0800_ErrorSignal.png}
%          \caption{Tracking error}
%          \label{fig:3harmError}
%      \end{subfigure}

%      \caption{Tracking performance and control signals for a reference signal with 3 harmonics.}
%      \label{fig:overall_tracking_results_3harm}
% \end{figure}

\section{Conclusion and Future Works}
In this paper, we presented a data-driven framework for designing feedforward steady-state generators for partially measured nonlinear systems with unknown dynamics. Motivated by the equivalence between time-domain and frequency-domain representations of steady-state inputs established in \cite{huang2001remarks}, the proposed method offers two main advantages over the classical model-based approach: it does not require prior knowledge of the plant model and avoids explicitly solving the regulator equations. In this sense, the method can be interpreted as a data-driven realization of the steady-state generator.

The framework was validated on two nonlinear systems commonly encountered in practice: a mechanical load with nonlinear friction driven by a DC motor, and an electrohydraulic actuator. For the friction example, the data-driven feedforward input was shown to recover the harmonic structure of the analytical solution to the regulator equations, providing direct empirical confirmation of the approach. For the electrohydraulic actuator, accurate tracking was demonstrated on multi-harmonic and square-wave references, illustrating the generalizability of the method to more complex nonlinear dynamics.

Despite these promising results, the proposed approach has two main limitations that will be addressed in future work. First, the amount of data required to train an accurate surrogate model grows with the number of harmonics retained, which can become prohibitive when high-bandwidth tracking is required; data-efficient training strategies and structured surrogate models are natural directions to address this. The second limitation is that the incorporation of feedback into the proposed feedforward design remains an open issue. Future work will focus on reducing the data requirements of the surrogate modeling stage and extending the proposed framework to include feedback for robustness.

\section{Acknowledgements}
Research was sponsored by the DEVCOM Army Research Laboratory and was accomplished under Cooperative Agreement Number W911NF-20-2-0161. The views and conclusions contained in this document are those of the authors and should not be interpreted as representing the official policies, either expressed or implied, of the DEVCOM Army Research Laboratory of the U.S. Government. The U.S. Government is authorized to reproduce and distribute reprints for Government.

\bibliographystyle{asmems4}
\bibliography{asme2e}.bib

\end{document}